\documentclass[letterpaper,10pt,twoside]{article}

\usepackage[utf8]{inputenc}  % xelatex does not need inputenc
\usepackage[english]{babel} % proper hyphenation support etc.

\usepackage{lmodern}
\usepackage{amssymb,amsthm,mathtools} % include AMS packages amssymb,mathtools,amsthm,amsmath
\usepackage{hyperref} % load this before geometry, autodetects pdftex/xetex
\usepackage{geometry} % letterpaper inherited from documentclass, autodetects pdftex/xetex

\usepackage{xcolor,graphicx}

\usepackage{pgfplots}
\pgfplotsset{compat=1.17}
\usepgfmodule{plot}
\usetikzlibrary{patterns}
\usetikzlibrary{calc}
\usetikzlibrary{decorations}
\usetikzlibrary{decorations.pathreplacing}
\usetikzlibrary{decorations.markings}

\newcommand{\im}{\mathrm{i}}

\newcommand{\R}{\mathbb{R}}
\newcommand{\bC}{\mathbb{C}}
\newcommand{\defeq}{\coloneqq}
\newcommand{\tens}{\otimes}
\DeclareMathOperator{\id}{id}
\newcommand{\xd}{\mathrm{d}}
\newcommand{\cH}{\mathcal{H}}
\newcommand{\cS}{\mathbf{S}}
\newcommand{\cA}{\mathbf{A}}
\newcommand{\po}{\mathsf{P}}
\newcommand{\op}{\mathcal{B}}
\newcommand{\coh}{\mathsf{K}}
\newcommand{\ncoh}{\mathsf{k}}

\allowdisplaybreaks

\begin{document}

%!TEX root = main.tex

\begin{titlepage}
\title{\textbf{Evanescent Particles}}
\author{%
  Daniele Colosi\footnote{email: dcolosi@enesmorelia.unam.mx}\\
  Escuela Nacional de Estudios Superiores, Unidad Morelia,\\
  Universidad Nacional Autónoma de México,\\
  C.P.~58190, Morelia, Michoacán, Mexico
  \and Robert Oeckl\footnote{email: robert@matmor.unam.mx}\\
  Centro de Ciencias Matemáticas,\\
  Universidad Nacional Autónoma de México,\\
  C.P.~58190, Morelia, Michoacán, Mexico}
\date{UNAM-CCM-2021-1\\ 25 April 2021\\ 10 November 2021 (v2)}

\maketitle

\vspace{\stretch{1}}

\begin{abstract}
Massive Klein-Gordon theory is quantized on a timelike hyperplane in Minkowski space using the framework of general boundary quantum field theory. In contrast to previous work, not only the propagating sector of the phase space is quantized, but also the evanescent sector, with the correct physical vacuum. This yields for the first time a description of the quanta of the evanescent field alone. The key tool is the novel $\alpha$-Kähler quantization prescription based on a $*$-twisted observable algebra. The spatial evolution of states between timelike hyperplanes is established and turns out to be non-unitary if different choices are made for the quantization ambiguity for initial and final hyperplane. Nevertheless, a consistent notion of transition probability is established also in the non-unitary case, thanks to the use of the positive formalism. Finally, it is shown how a conducting boundary condition on the timelike hyperplane gives rise to what we call the Casimir state. This is a pseudo-state which can be interpreted as an alternative vacuum and which gives rise to a sea of particle pairs even in this static case.
\end{abstract}

\vspace{\stretch{1}}
\end{titlepage}

%\tableofcontents

% !TEX root = main.tex

\section{Introduction}

Evanescent waves play an important role in many electromagnetic phenomena. These can often be successfully described by classical electrodynamics. When it becomes necessary to go beyond a classical treatment, the problem of the quantization of evanescent waves arises. In free field theory in vacuum Minkowski space, well-behaved global solutions are described by propagating waves. Consequently, these form the exclusive basis of traditional quantization prescriptions and thus of quantum field theory. In a seminal work almost 50 years ago, Carniglia and Mandel performed a quantization of evanescent waves in free electrodynamics \cite{CarMan:quantevanemw}. This was achieved by introducing a medium in half of space with a refractive index larger than one and quantizing the classical solutions in the conventional way. These solutions consist of propagating waves incident, transmitted or reflected on the interface separating the two regions of space, as well as evanescent waves near the interface. This work has formed the basis for subsequent advances in the quantization of evanescent waves \cite{BeSiGu:evanwoptics}. Unfortunately, in this way it is not possible to isolate the quantum excitations associated to evanescent waves alone, as they always appear as parts of solutions that involve propagating waves as well.

In order to describe the quanta that correspond exclusively to evanescent waves we need to break away from the traditional global approach to quantization which is the root cause for our difficulties in isolating them. Thanks to recent advances in the foundations of quantum field theory, it is now possible to quantize classical field theory degrees of freedom on general hypersurfaces in spacetime, not only spacelike ones. 
The relevant phase space in this case is the space of \emph{germs of solutions} on the hypersurface. Informally speaking, this is the space of solutions defined near the hypersurface. If the hypersurface is spacelike, these solutions correspond to global solutions, as usual. If the hypersurface is not spacelike, these solutions need not be defined in all of spacetime. Crucially, the phase space for a timelike hypersurface generally contains evanescent waves in addition to propagating ones.
For an exposition of the underlying theoretical framework and its development we refer the reader to \cite{Oe:boundary,Oe:gbqft,Oe:probgbf,Oe:reveng,Oe:posfound} and references therein.

We shall consider in the following the simplest possible situation: An infinitely extended plane surface in space, extended over all of time. We refer to this as the \emph{timelike hyperplane}. Also, in order to avoid inessential complications due to gauge degrees of freedom, we shall not consider electrodynamics, but the free scalar field, that is, \emph{Klein-Gordon theory}.
The first field-theoretic quantization on a timelike hypersurface was proposed and carried out about 15 years ago, precisely in the context of Klein-Gordon theory on the timelike hyperplane \cite{Oe:timelike}. The construction of the Hilbert space of states, based on the Schrödinger representation, works surprisingly well and paved the way for many posterior developments. However, it has one crucial flaw. It is limited to the propagating sector of the phase space and excludes the evanescent modes.
In the free theory without sources this works fine \cite{Oe:kgtl} since the propagating and evanescent modes decouple. However, once sources or interactions are introduced, the evanescent modes need to be taken into account. In an asymptotic setting it is possible (and indeed physically correct) to exclude them from the state space \cite{CoOe:spsmatrix,CoOe:smatrixgbf}. However, at finite location of the hypersurface they must be included.

A technical reason for the exclusion of the evanescent modes in the work \cite{Oe:timelike} was the difficulty of constructing a Schrödinger representation for them. In contrast, the propagating modes on the timelike hyperplane admit a Schrödinger quantization quite analogous to the usual one on a spacelike hypersurface. Using the holomorphic representation, a proposal for a quantization on the timelike hyperplane that includes the evanescent modes was put forward in \cite{Oe:holomorphic}. This was accomplished by mimicking the canonical quantization prescription for the propagating modes in terms of postulating an ad hoc complex structure. However, the choice of complex structure was solely guided by positive-definiteness and analytical continuation, but lacked a physical justification.

The root cause for the difficulties in the quantization of evanescent modes was understood only recently \cite{CoOe:vaclag}. In standard canonical quantization on a spacelike hypersurfaces there is a choice of vacuum encoded in a complex structure on the phase space. This choice determines an inner product, the Fock space, etc. The choice of vacuum originates from a choice of Wick-rotated asymptotic boundary conditions of the field. The latter give rise to a Kähler polarization which in turn gives rise to a complex structure and thus a canonical quantization. For propagating modes on a timelike hypersurface this works exactly the same way, even though the boundary condition resides now at spatial rather than temporal infinity. In contrast, for evanescent modes, the correct physical asymptotic boundary conditions are decaying conditions, which are real and not Wick-rotated. They give rise to a real polarization which is not Kähler, do not induce a complex structure and hence do not lead to a canonical quantization.

In the meantime, a new quantization prescription has been developed precisely to address vacua corresponding to non-Kähler polarizations \cite{CoOe:locgenvac}. Based on this, in the present work we are finally able to present a fully satisfactory quantization of Klein-Gordon theory on the timelike hyperplane that includes the evanescent modes. We start with a brief review of canonical quantization on equal-time hypersurfaces to fix notation and introduce the appropriate ingredients in Section~\ref{sec:slquant}. In Section~\ref{sec:Hstml} we present the construction of the Hilbert space on the timelike hyperplane, the representation of observables and coherent states. The \emph{spatial} evolution between timelike hyperplanes is the subject of Section~\ref{sec:spevol}. The associated transition probabilities are worked out in Section~\ref{sec:transprob}. In Section~\ref{sec:casimir} we take advantage of the framework developed in \cite{CoOe:locgenvac} to implement the reflecting boundary condition of the Casimir effect as a pseudo-state on the timelike hyperplane encoding a change of vacuum. Finally, we present conclusions and an outlook is Section~\ref{sec:conclusions}. We use throughout tools and conventions from General Boundary Quantum Field Theory (GBQFT), see \cite{Oe:gbqft,Oe:holomorphic,Oe:feynobs,CoOe:locgenvac} and references therein, but try to keep this paper as self-contained as possible. When considering predictions and probabilities we resort to the positive formalism where necessary, see \cite{Oe:posfound} and references therein.

% !TEX root = main.tex

\section{The Hilbert space on a spacelike hyperplane reviewed}
\label{sec:slquant}

To fix notation and clarify the basic concepts, we consider the massive Klein-Gordon theory in Minkowski space first in the familiar and long-established setting of canonical quantization on equal-time hypersurfaces. We follow the conventions of \cite{Oe:holomorphic}. Fix a time $t$ and denote the space of germs of solutions of the Klein-Gordon equation at $t$ by $L_t$. Alternatively, $L_t$ is the \emph{phase space} or \emph{space of initial data} at time $t$. Also, this space is in one-to-one correspondence to the space of global solutions, due to the Cauchy property. We parametrize \emph{complexified} solutions, i.e., elements of $L_t^\bC$, as usual in terms of plane waves, with $E=\sqrt{k^2+m^2}$,
\begin{equation}
 \phi(t,x)=\int\frac{\xd^3 k}{(2\pi)^3 2E}
 \left(\phi^\text{a}(k) e^{-\im(E t-k x)}+\overline{\phi^\text{b}(k)} e^{\im(E t-k x)}\right) .
 \label{eq:kgmodes}
\end{equation}
Real solutions, i.e., elements of $L_t$ are characterized by the property $\phi^\text{b}(k)=\phi^\text{a}(k)$. Important ingredients for encoding the classical dynamics on $L_t$ are the bilinear \emph{symplectic potential} $[\cdot,\cdot]_t:L_t\times L_t\to\R$ and the anti-symmetric bilinear \emph{symplectic form} $\omega_t:L_t\times L_t\to\R$ given by,
\begin{align}
  [\phi,\eta]_t & =\int\xd^3 x\, \eta(t,x) \partial_0 \phi(t,x) \nonumber \\
  & =\frac{\im}{2}\int\frac{\xd^3 k}{(2\pi)^3 2E}
  \left(\eta^\text{a}(k)\overline{\phi^\text{b}(k)}-\phi^\text{a}(k)\overline{\eta^\text{b}(k)}
  -\phi^\text{a}(k)\eta^\text{a}(-k) e^{-2\im E t}
  +\overline{\phi^\text{b}(k)}\overline{\eta^\text{b}(-k)} e^{2\im E t}\right) \label{eq:spkgm} \\
  \omega_t(\phi,\eta)  & =\frac{1}{2}\left([\phi,\eta]_t-[\eta,\phi]_t\right) \nonumber \\
  & =\frac{\im}{2}\int\frac{\xd^3 k}{(2\pi)^3 2E}
  \left(\eta^\text{a}(k)\overline{\phi^\text{b}(k)}-\phi^\text{a}(k)\overline{\eta^\text{b}(k)}\right) .
  \label{eq:sfkgm}
\end{align}
Recall that the symplectic potential and symplectic form change sign under a change of orientation of the hypersurface \cite{Oe:holomorphic}, here the equal-time hypersurface at time $t$. This is of little relevance in conventional treatments of canonical quantization where all spacelike hypersurfaces are implicitly equipped with the same orientation and the issue of signs reduces to a single global choice of sign. In GBQFT on the other hand, hypersurfaces inherit an orientation from the region they bound. In particular, an initial and a final spacelike boundary hypersurface for a time-interval region carry opposite orientations. The orientation chosen for formulas (\ref{eq:spkgm}) and (\ref{eq:sfkgm}) to carry the correct sign is that of an initial hypersurface.\footnote{We follow the sign conventions in \cite{Oe:holomorphic,CoOe:locgenvac}, opposite to those in \cite{CoOe:vaclag}.}
This is in accordance with the requirements for constructing the Hilbert space $\cH_t$ of \emph{initial} states, that is the “ket-states”, at time $t$.

We follow \cite{CoOe:vaclag} in formulating the subsequent steps in constructing this Hilbert space $\cH_t$. To this end consider the \emph{inner product} on $L_t^\bC$, i.e., the hermitean sesquilinear form given by,
\begin{equation}
  \left(\phi,\eta\right)_t\defeq 4\im\omega_t\left(\overline{\phi},\eta\right) .
  \label{eq:stdipc}
\end{equation}
Recall also that a \emph{Lagrangian subspace} $L_t^+$ of the symplectic vector space $L_t^\bC$ is characterized by the following properties. $L_t^+$ is \emph{isotropic}, i.e.,
\begin{align}
  \omega_{t}(\phi,\eta) & =0,\quad\forall \phi,\eta\in L_t^{+},\\
  \text{as well as \emph{coisotropic}},\quad \omega_{t}(\phi,\eta) & =0,\quad\forall\phi\in L_t^+ \Rightarrow \eta\in L_t^+ .
\end{align}

Construction of the Hilbert space $\cH_t$ of states requires the choice of a \emph{vacuum}. In standard canonical quantization this is encoded in a Lagrangian subspace $L_t^+\subseteq L_t^\bC$ on which the inner product (\ref{eq:stdipc}) is \emph{positive-definite}. We refer to a choice of positive-definite Lagrangian subspace as a \emph{Kähler polarization}, a nomenclature coming from geometric quantization \cite{Woo:geomquant}. The Kähler polarization encoding a vacuum is usually determined by Wick-rotated asymptotic boundary conditions of the field \cite{CoOe:vaclag}. In the case at hand, the (past) vacuum corresponds to the positive-definite Lagrangian subspace of \emph{negative energy} solutions,
\begin{equation}
  L^+_t=\{\phi\in L_t^\bC : \phi^\text{a}(k)=0\,\forall k\} .
  \label{eq:kgpvac}
\end{equation}
The inner product (\ref{eq:stdipc}) on $L_t^+$ takes the form,
\begin{equation}
  (\phi,\eta)_t  = 2\int\frac{\xd^3 k}{(2\pi)^3 2E}
   \phi^\text{b}(k)\overline{\eta^\text{b}(k)} .
\end{equation}
 
In canonical quantization the vacuum on the other side of the hypersurface, i.e., here to the future of $t$, is given by the complex conjugate subspace $L_t^-\defeq \overline{L_t^+}$. This is then automatically a positive-definite Lagrangian subspace with respect to the hypersurface with opposite orientation, i.e., with the symplectic form~(\ref{eq:sfkgm}) taking the opposite sign. Here, $L_t^-$ is the subspace of \emph{positive energy} solutions,
\begin{equation}
  L^-_t=\{\phi\in L_t^\bC : \phi^\text{b}(k)=0\,\forall k\} .
%  \label{eq:kgnvac}
\end{equation}

A Kähler polarization can be encoded equivalently in terms of a \emph{complex structure}, i.e., a complex linear map $J_t:L_t^\bC\to L_t^\bC$ such that $J_t^2=-\id_t$ and which is compatible with the symplectic form in the sense, $\omega_t(J_t \phi, J_t\phi')=\omega_t(\phi,\phi')$. The Kähler polarization is recovered from the complex structure in terms of the eigenspaces for the eigenvalues $\im$ and $-\im$, which are precisely $L_t^+$ and $L_t^-$. Here, the complex structure corresponding to the Kähler polarization of the standard vacuum is given by,
\begin{equation}
  (J_t(\phi))^{\text{a/b}}(k) =-\im \phi^{\text{a/b}}(k) .
\end{equation}
We recall that a complex structure corresponding to a Kähler polarization also gives rise to a positive-definite complex inner product on the real phase space $L_t$. This is the restriction to $L_t$ of the bilinear form on $L_t^{\bC}$ given by, 
\begin{equation}
  \{\phi,\eta\}_t\defeq 2\omega_t(\phi,J_t \eta)+2\im\omega_t(\phi,\eta)
  =4\im\omega_t(\phi^-,\eta^+) .
  \label{eq:rjip}
\end{equation}
Here, the notation $\xi=\xi^+ +\xi^-$ is used for the decomposition $L^\bC_t=L^+_t \oplus L^-_t$.
This is in one-to-one correspondence to the inner product (\ref{eq:stdipc}) on $L_t^+$ via the identification $L_t\to L_t^+$ given by $\phi\mapsto \phi^+$. Here we have,
\begin{equation}
  \{\phi,\eta\}_t = 2\int\frac{\xd^3 k}{(2\pi)^3 2E}
  \phi^\text{a}(k)\overline{\eta^\text{b}(k)} .
\end{equation}
The creation and annihilation operators are parametrized by elements of $L_t$ and their commutation relations are determined by this same inner product,
\begin{equation}
  [a_{\eta},a_{\phi}^{\dagger}]=\{\phi,\eta\}_t .
  \label{eq:ccr}
\end{equation}

A particularly important class of states that we shall make extensive use of are the \emph{coherent states}. These are obtained by acting with exponentiated creation operators on the vacuum state and generate a dense subspace of the Hilbert space, see e.g., \cite{ItZu:qft}. A coherent state $\coh_\xi\in\cH_t$ is labeled by an element of the phase space $\xi\in L_t$ and in many respects behaves as an approximation of this classical phase space element. Explicitly,
\begin{equation}
  K_{\xi}=\exp\left(\frac{1}{\sqrt{2}} a^{\dagger}_\xi\right) K_0,
\end{equation}
where $K_0$ is the vacuum state.
The inner product of coherent states is given by,
\begin{equation}
  \langle \coh_\xi,\coh_\phi\rangle_t =\exp\left(\frac12\{\phi,\xi\}_t\right) .
  \label{eq:cohip}
\end{equation}
The coherent states satisfy the completeness relation,
\begin{equation}
  \langle\eta,\psi\rangle_{t} = \int_{\hat{L}_{t}} \langle \eta, \coh_{\phi}\rangle_{t} \langle \coh_{\phi}, \psi\rangle_{t}\,\xd\nu_{t}(\phi) .
  \label{eq:cohcompl}
\end{equation}
Here, the integral is over an extension $\hat{L}_{t}$ of the phase space $L_t$ with the Gaussian measure $\nu_t$, determined by the inner product on $L_t$ \cite{Oe:holomorphic}.
In the following we will also consider \emph{normalized} coherent states, given by,
\begin{equation}
  \ncoh_\xi=\exp\left(-\frac14 \{\xi,\xi\}_t\right) \coh_\xi .
\end{equation}

We proceed to discuss the representation of observables on the Hilbert space $\cH_t$ of states. We can think of observables here as functions on the instantaneous phase space $L_t$, although they should really be thought of as arising from \emph{slice observables} in spacetime \cite{Oe:feynobs,CoOe:locgenvac}. What is more, we require them to extend to holomorphic functions on the complexified phase space $L_t^\bC$ \cite{CoOe:locgenvac}. We restrict our considerations to \emph{Weyl observables}, which arise as follows. Associated to $\xi\in L_t^\bC$ we define the linear observable $D_\xi:L_t^\bC\to\bC$ and the Weyl observable $F_\xi:L_t^\bC\to\bC$ given by,
\begin{equation}
  D_\xi(\phi)\defeq 2\omega_t(\xi,\phi), \qquad
  F_\xi\defeq \exp(\im D) .
\end{equation}
The action of the quantized Weyl observable $\hat{F}_\xi$ on a coherent state $\coh_\phi$ is \cite{CoOe:locgenvac},
\begin{equation}
  \hat{F}_\xi \coh_\phi=\exp\left(-\frac12 \{\phi,\xi\}_t-\frac14 \{\xi,\xi\}_t\right) \coh_{\phi+\xi^- +\overline{\xi^-}} .
  \label{eq:weylactcoh}
\end{equation}
In particular, we see that the quantized Weyl observables satisfy the \emph{Weyl relations},
\begin{equation}
  \hat{F}_\xi \hat{F}_\phi=\exp(\im\omega_t(\xi,\phi))\hat{F}_{\xi+\phi} .
  \label{eq:weylrel}
\end{equation}
If $\xi$ is real, i.e., $\xi\in L_t$ so that $D_\xi$ is also real, the action of $\hat{F}_\xi$ becomes unitary. On normalized coherent states we then have,
\begin{equation}
  \hat{F_\xi} \ncoh_\phi=\exp\left(\im\omega_t(\xi,\phi)\right) \ncoh_{\phi+\xi} .
  \label{eq:weylactunitary}
\end{equation}

\section{The Hilbert space on a timelike hyperplane}
\label{sec:Hstml}

\begin{figure}
  \centering
  \begin{tikzpicture}[scale=1]
\filldraw[color=gray!10!] (2,-0.5) rectangle(5,5);
\draw[thick] (2,-0.5) node [below] {$\Sigma$} -- (2,5);
\draw[->] (-0.5,0) -- (5,0) node [right] {$x$};
\draw[->] (0,-0.5) -- (0,5) node [left] {$t$};
\draw[->] (1.6,4.75) -- (0.6,4.75);  
\draw[->] (1.6,3.75) -- (0.6,3.75);  
\draw[->] (1.6,2.75) -- (0.6,2.75);  
\draw[->] (1.6,1.75) -- (0.6,1.75);  
\draw[->] (1.6,.75) -- (0.6,.75);  
\draw[->] (1.6,-0.25) -- (0.6,-0.25);  
\node at (1,2.2) {vacuum};
\node at (1,1.2) {boundary};
\node at (1,0.28) {condition};
\end{tikzpicture}
  \caption{The timelike hyperplane $\Sigma$ in Minkowski space.}
  \label{fig:tlhp}
\end{figure}
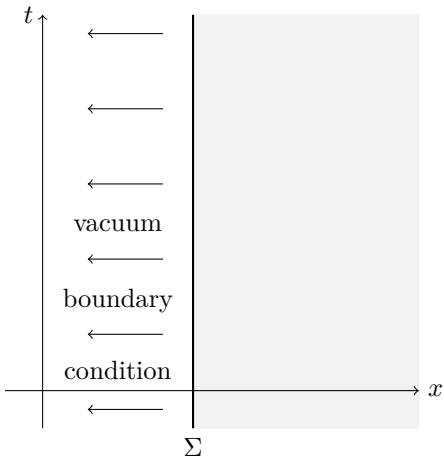

We consider a timelike hyperplane in Minkowski space characterized by a fixed value of the coordinate $x_1$, say at $x_1=z$, see Figure~\ref{fig:tlhp}. We wish to construct the space of states of massive Klein-Gordon quantum field theory on the hypersurface. To be specific about the orientation, we consider the hypersurface as the boundary of the half space $x_1\ge z$. The first step is to consider the space $L_z$ of germs of solutions on the hypersurface. This naturally splits into a direct sum of the subspaces of \emph{propagating solutions} and of \emph{evanescent solutions}, $L_z=L_z^{\text{p}}\oplus L_z^{\text{e}}$.
In the following we write $\tilde{x}=(x_2,x_3)$ and $\tilde{k}=(k_2,k_3)$ as a collective notation for positions and momenta in the two spatial coordinate directions tangential to the hypersurface.
The space $L_z^{\text{p},\bC}$ of complexified propagating solutions may be parametrized as,
\begin{equation}
\phi(t,x_1,\tilde{x}) =  \int_{|E|>E_\parallel} \frac{\xd^2\tilde{k}\, \xd E}{(2 \pi)^{3} 2k_1}\,
   e^{-\im E t + \im \tilde{k} \tilde{x}} \left( \phi^{\text{a}}(E, \tilde{k})  e^{\im k_1 x_1} + \phi^{\text{b}}(E, \tilde{k})  e^{-\im k_1 x_1} \right),
\label{eq:kgtlp}
\end{equation}
where $E_\parallel=\sqrt{\tilde{k}^2+m^2}$ and $k_1 = \sqrt{|E^2-\tilde{k}^2-m^2|}$. The integral in $\tilde{k}$ is over $\R^2$. Here, $\phi^{\text{a}},\phi^{\text{b}}$ are complex functions on the subset $\{(E,\tilde{k})\in\R\times \R^2: |E|>E_\parallel\}$.
These solutions are characterized by an oscillatory behavior in the $x_1$ direction. They are precisely the usual plane wave solutions as for the equal-time hypersurface (\ref{eq:kgmodes}), only slightly differently parametrized.
We parametrize the space $L_z^{\text{e},\bC}$ of complexified evanescent solutions as,
\begin{equation}
\phi(t,x_1,\tilde{x}) = \int_{|E|< E_\parallel} \frac{\xd^2\tilde{k}\, \xd E }{(2 \pi)^{3} 2k_1}\, e^{-\im E t + \im \tilde{k} \tilde{x}} \left( \phi^\text{x}(E, \tilde{k})  e^{- k_1 x_1} + \phi^\text{i}(E, \tilde{k})  e^{ k_1 x_1} \right) .
\label{eq:kgtle}
\end{equation}
Here, $\phi^{\text{x}},\phi^{\text{i}}$ are complex functions on the subset $\{(E,\tilde{k})\in\R\times \R^2: |E|<E_\parallel\}$.
This space consists of solutions that grow or decay exponentially in the direction perpendicular to the hypersurface. Note that the parametrizations~(\ref{eq:kgtlp}) and (\ref{eq:kgtle}) are global in the sense of being independent of the location $z$ of the hyperplane. Consequently, the spaces $L_z$ for different $z$ are really the same. Sometimes we shall write $L$ instead of $L_z$ to emphasize this. Similarly, we shall sometimes leave out the subscript $z$ for structures associated to $L_z$ when they do not depend on the choice of $z$.

The symplectic potential on the hypersurface (as a boundary of a region with $x_1\ge z$) is the bilinear form
$L_z \times L_z \to \R$  given by
\begin{equation}
  [\phi,\eta]_z =  - \int \xd t \, \xd^2\tilde{x}\, \eta(t,z,\tilde{x}) (\partial_1 \phi) (t,z,\tilde{x}) .
\end{equation}
Here, $\partial_1$ denotes the partial derivative corresponding to the coordinate $x_1$. That is, it is the normal derivative to the hyperplane.
We exhibit the propagating and evanescent contributions to the symplectic potential separately,
\begin{align}
  [\phi,\eta]^{\text{p}}_z & = \frac{\im}{2}\int_{|E|>E_\parallel}
  \frac{\xd^2\tilde{k}\,\xd E}{(2\pi)^3 2k_1}
  \left(\eta^\text{a}(E,\tilde{k})\phi^\text{b}(-E,-\tilde{k})
  -\eta^\text{b}(-E,-\tilde{k})\phi^\text{a}(E,\tilde{k}) \right . \nonumber \\
  & \quad \left . 
  -\eta^\text{a}(E,\tilde{k})\phi^\text{a}(-E,-\tilde{k}) e^{2\im k_1 z}
  +\eta^\text{b}(E,\tilde{k})\phi^\text{b}(-E,-\tilde{k}) e^{-2\im k_1 z}
  \right) ,
  \label{eq:tlsympotp} \\
  [\phi,\eta]^{\text{e}}_z & = -\frac{1}{2}\int_{|E|< E_\parallel}
  \frac{\xd^2\tilde{k}\,\xd E}{(2\pi)^3 2k_1}
  \left(\eta^\text{x}(E,\tilde{k})\phi^\text{i}(-E,-\tilde{k})
  -\eta^\text{i}(-E,-\tilde{k})\phi^\text{x}(E,\tilde{k}) \right . \nonumber \\
  & \quad \left . 
  -\eta^\text{x}(E,\tilde{k})\phi^\text{x}(-E,-\tilde{k}) e^{-2k_1 z}
  +\eta^\text{i}(E,\tilde{k})\phi^\text{i}(-E,-\tilde{k}) e^{2k_1 z}
  \right) .
  \label{eq:tlsympote}
\end{align}
The symplectic form $\omega:L \times L \to \R$ is the anti-symmetric part of the symplectic potential and the standard inner product is given by (\ref{eq:stdipc}). Again, we exhibit the propagating and evanescent contributions separately,
\begin{align}
  \omega^{\text{p}}(\phi,\eta) & = \frac{\im}{2}\int_{|E|>E_\parallel}
  \frac{\xd^2\tilde{k}\,\xd E}{(2\pi)^3 2k_1}
  \left(\eta^\text{a}(E,\tilde{k})\phi^\text{b}(-E,-\tilde{k})
  -\eta^\text{b}(-E,-\tilde{k})\phi^\text{a}(E,\tilde{k})\right) ,
  \label{eq:tlsymfp} \\
  \omega^{\text{e}}(\phi,\eta) & = -\frac{1}{2}\int_{|E|< E_\parallel}
  \frac{\xd^2\tilde{k}\,\xd E}{(2\pi)^3 2k_1}
  \left(\eta^\text{x}(E,\tilde{k})\phi^\text{i}(-E,-\tilde{k})
  -\eta^\text{i}(-E,-\tilde{k})\phi^\text{x}(E,\tilde{k})\right) ,
  \label{eq:tlsymfe} \\
  (\phi,\eta)^{\text{p}} & = 2\int_{|E|>E_\parallel}
  \frac{\xd^2\tilde{k}\,\xd E}{(2\pi)^3 2k_1}
  \left(\eta^\text{b}(E,\tilde{k})\overline{\phi^\text{b}(E,\tilde{k})}-\eta^\text{a}(E,\tilde{k})\overline{\phi^\text{a}(E,\tilde{k})}\right) ,
  \label{eq:tlipp} \\
  (\phi,\eta)^{\text{e}} & = 2\im\int_{|E|<E_\parallel}
  \frac{\xd^2\tilde{k}\,\xd E}{(2\pi)^3 2k_1}
  \left(\eta^\text{i}(E,\tilde{k})\overline{\phi^\text{x}(E,\tilde{k})}-\eta^\text{x}(E,\tilde{k})\overline{\phi^\text{i}(E,\tilde{k})}\right) .
  \label{eq:tlipe}
\end{align}

The vacuum in terms of the asymptotic boundary condition to the left (compare Figure~\ref{fig:tlhp}) is given for the propagating modes by the subspace \cite{CoOe:vaclag},
\begin{equation}
  L^{\text{p},+}=\{\phi\in L^{\text{p},\bC}:\;\phi^{\text{a}}(E,\tilde{k})=0\;\forall E,\tilde{k}\} .
  \label{eq:tlpvac}
\end{equation}
(Recall that our orientation conventions here are opposite to those of the cited paper.) This is easily seen to be a Lagrangian subspace. Moreover, it is a positive-definite subspace with respect to the inner product (\ref{eq:tlipp}). This allows us to perform the quantization in analogy to the spacelike hypersurface case of Section~\ref{sec:slquant}, yielding a Hilbert space of states that we shall denote $\cH^{\text{p}}$. This was previously described in \cite{Oe:holomorphic}, but is also equivalent to the Schrödinger quantization first carried out for this hypersurface in \cite{Oe:timelike}. For completeness, we mention that the vacuum to the right is determined by the complex conjugate subspace $L^{\text{p},-}=\overline{L^{\text{p},+}}$ which encodes the asymptotic boundary conditions to the right,
\begin{equation}
  L^{\text{p},-}=\{\phi\in L^{\text{p},\bC}:\;\phi^{\text{b}}(E,\tilde{k})=0\;\forall E,\tilde{k}\} .
%  \label{eq:tlpvacr}
\end{equation}
We also note that the bilinear form (\ref{eq:rjip}) is given here by,
\begin{equation}
  \{\phi,\eta\}^{\text{p}}= 2\int_{|E|>E_\parallel}
  \frac{\xd^2\tilde{k}\,\xd E}{(2\pi)^3 2k_1}
  \eta^\text{b}(-E,-\tilde{k})\phi^\text{a}(E,\tilde{k}) ,
\end{equation}
giving rise to a positive-definite complex inner product on $L^{\text{p}}$. As in the spacelike case, this also determines the commutation relations between creation and annihilation operators via the analogue of (\ref{eq:ccr}).

For the evanescent modes the physically correct vacuum boundary condition to the left is a decay condition, yielding the subspace \cite{CoOe:vaclag},
\begin{equation}
  L^{\text{e},+} = \{\phi\in L^{\text{e},\bC}:\; \phi^{\text{x}}(E,\tilde{k})=0\;\forall E,\tilde{k}\} .
  \label{eq:tlevac}
\end{equation}
Again, $L^{\text{e},+}$ is a Lagrangian subspace of $L^{\text{e},\bC}$. However, it is a \emph{neutral} rather than positive-definite subspace with respect to the inner product (\ref{eq:tlipe}). Correspondingly, it is a \emph{real Lagrangian subspace}, i.e., it arises as the complexification of a real subspace. Similarly, the vacuum boundary condition to the right is a decay condition to the right,
\begin{equation}
  L^{\text{e},-} = \{\phi\in L^{\text{e},\bC}:\; \phi^{\text{i}}(E,\tilde{k})=0\;\forall E,\tilde{k}\} .
%  \label{eq:tlevacr}
\end{equation}
This is also a real Lagrangian subspace of $L^{\text{e},\bC}$. With the polarizations determined, we write down the bilinear form given by formula (\ref{eq:rjip}),
\begin{equation}
  \{\phi,\eta\}^{\text{e}}  = 2\im\int_{|E|< E_\parallel}
  \frac{\xd^2\tilde{k}\,\xd E}{(2\pi)^3 2k_1}
  \eta^\text{i}(-E,-\tilde{k})\phi^\text{x}(E,\tilde{k}) .
  \label{eq:rjipe}
\end{equation}

With the inner product (\ref{eq:tlipe}) not positive-definite on $L^{\text{e},+}$, canonical quantization fails. Instead, we shall use the novel methods of \cite{CoOe:locgenvac} to construct a Hilbert space for the evanescent modes. This requires an additional ingredient, a \emph{compatible real structure} $\alpha:L^{\text{e},\bC}\to L^{\text{e},\bC}$. Being a real structure, $\alpha$ behaves as a complex conjugation, i.e., it is complex conjugate linear, $\alpha(\lambda\phi)=\overline{\lambda}\alpha(\phi)$ for $\lambda\in\bC$ and $\alpha(\alpha(\phi))=\phi$. Compatibility refers to compatibility with the symplectic form in the sense,
\begin{equation}
  \omega^{\text{e}}(\alpha(\phi),\alpha(\eta))=\overline{\omega^{\text{e}}(\phi,\eta)} .
  \label{eq:alphacomp}
\end{equation}
Lastly, we require \emph{positive definiteness} of the following \emph{inner product},
\begin{equation}
  (\phi,\eta)^{\text{e},\alpha}\defeq 4\im\omega^{\text{e}}(\alpha(\phi),\eta) ,
  \label{eq:tlipea}
\end{equation}
on the polarized subspace $L^{\text{e},+}\subseteq L^{\text{e},\bC}$. This allows the construction of a Hilbert space $\cH^{\text{e},\alpha}$ by $\alpha$-Kähler quantization, a method quite analogous to the usual canonical (Kähler) quantization, but where crucially the inner product (\ref{eq:tlipe}) is replaced by the inner product (\ref{eq:tlipea}). We may also think of $\alpha$-Kähler quantization as amounting to the replacement of the real subspace $L\subseteq L^{\bC}$ by the real subspace $L^{\alpha}\subseteq L^{\bC}$, where $L^{\alpha}\defeq \{\phi\in L^{\bC}: \alpha(\phi)=\phi\}$. In particular, the bilinear form (\ref{eq:rjipe}) becomes a complex inner product on $L^{\alpha}$ with the complex structure $J$ determined by taking eigenvalues $+\im$ and $-\im$ on the polarized subspaces $L^+$ and $L^-$ respectively. Creation and annihilation operators are then also parametrized by elements of $L^{\alpha}$ with commutation relations given as in (\ref{eq:ccr}), but with the inner product (\ref{eq:rjipe}).

The question remains how we obtain a compatible real structure in the present case. As the polarization~(\ref{eq:tlevac}) encoding the vacuum is real, we may obtain such a structure from a \emph{positive-definite reflection map} \cite[Section~7.7]{CoOe:locgenvac}. It turns out that in the present case such a map arises from an actual reflection in spacetime, quite as the name suggests. Concretely, consider the effect of a reflection at the timelike hyperplane with $x_1=u$ on solutions,
\begin{equation}
  (\gamma_u(\phi))(t,x_1,\tilde{x})=\phi(t,2u-x_1,\tilde{x}) .
  \label{eq:reflects}
\end{equation}
We can think of $\gamma_u$ as an operator on the solution space $L_z^{\bC}$, where $z$ does not need to coincide with $u$. In terms of the mode expansions (\ref{eq:kgtlp}) and (\ref{eq:kgtle}) this takes the form,
\begin{align}
  (\gamma_u(\phi))^{\text{b}}(E,\tilde{k})
   & =e^{2\im k_1 u}\phi^{\text{a}}(E,\tilde{k}), \qquad
  (\gamma_u(\phi))^{\text{a}}(E,\tilde{k})
    =e^{-2\im k_1 u}\phi^{\text{b}}(E,\tilde{k}) , \\
  (\gamma_u(\phi))^{\text{x}}(E,\tilde{k})
  & =e^{2k_1 u}\phi^{\text{i}}(E,\tilde{k}), \qquad
  (\gamma_u(\phi))^{\text{i}}(E,\tilde{k})
    =e^{-2k_1 u}\phi^{\text{x}}(E,\tilde{k}) .
\end{align}
We read off immediately that $\gamma_u$ interchanges the polarizations corresponding to left and right vacuum boundary conditions. For the propagating modes we have no further need for this structure, but for the evanescent modes we define $\alpha^{\text{e}}_u(\phi)\defeq -\im\overline{\gamma_u(\phi)}$ for $\phi\in L^{\text{e},\bC}$.  We may then verify that $\alpha^{\text{e}}_u$ is indeed a compatible real structure. Explicitly,
\begin{gather}
  (\alpha_u^{\text{e}}(\phi))^{\text{x}}(E,\tilde{k})
  =-\im e^{2k_1 u}\overline{\phi^{\text{i}}(-E,-\tilde{k})}, \qquad
  (\alpha_u^{\text{e}}(\phi))^{\text{i}}(E,\tilde{k})
  =-\im e^{-2k_1 u}\overline{\phi^{\text{x}}(-E,-\tilde{k})} , \\
  (\phi,\eta)^{\text{e},u} = 2\int_{|E|<E_\parallel}
  \frac{\xd^2\tilde{k}\,\xd E}{(2\pi)^3 2k_1}
  \left(e^{2 k_1 u}\eta^\text{i}(E,\tilde{k})\overline{\phi^\text{i}(E,\tilde{k})}-e^{-2 k_1 u}\eta^\text{x}(E,\tilde{k})\overline{\phi^\text{x}(E,\tilde{k})}\right) , \\
  L^{\text{e},u}= \left\{\phi\in L^{\text{e},\bC}: \phi^{\text{x}}(E,\tilde{k})=-\im e^{2k_1 u}\overline{\phi^{\text{i}}(-E,-\tilde{k})}\right\} .
  \label{eq:Les}
\end{gather}
In the last two lines we have used a simplified notation by replacing the superscripts $\alpha^{\text{e}}_u$ by $u$. As mentioned previously, we can now perform $\alpha$-Kähler quantization to obtain a Hilbert space of states for the evanescent modes at the $x_1=z$ hyperplane. Since the construction does not depend explicitly on the position $z$ of the hyperplane, but does depend on $u$ we use the notation $\cH^{\text{e},u}$ for this Hilbert space. The total Hilbert space of states for the hypersurface $x_1=z$ is the (completed) tensor product $\cH^u=\cH^{\text{p}}\tens\cH^{\text{e},u}$. Again, as the construction does not explicitly depend on the hypersurface position $z$, we omit a corresponding subscript in our notation. Note that instead of constructing the components for propagating and evanescent solutions separately, we could have constructed the total Hilbert space in a single step via $\alpha$-Kähler quantization. The compatible real structure $\alpha_u:L^{\bC}\to L^{\bC}$ is given by $\alpha_u(\phi)=\alpha^{\text{e}}_u(\phi^{\text{e}})+\overline{\phi^\text{p}}$, where $\phi^{\text{e}}$ and $\phi^{\text{p}}$ denote here the evanescent and propagating components of $\phi\in L^{\bC}=L^{\text{e},\bC}\oplus L^{\text{p},\bC}$ respectively.

As for coherent states, their construction and properties on the Hilbert space $\cH^\text{p}$ of propagating modes is completely analogous to the standard case of spacelike hypersurfaces. What is more, we have shown in \cite{CoOe:locgenvac} that there is also a straightforward generalization in $\alpha$-Kähler quantization, which applies here to the Hilbert space $\cH^{\text{e},u}$ of evanescent modes. Coherent states are still obtained through the action of exponentiated creation operators on the vacuum state. However, creation operators and consequently coherent states are now labeled by elements of $L^\alpha$ rather than the real phase space $L$. In the present context, coherent states on $\cH^u$ are thus labeled by elements of $L^u=L^{\text{p}}\oplus L^{\text{e},u}$. The inner product of coherent states resembles that (\ref{eq:cohip}) of the standard case,
\begin{equation}
  \langle \coh^u_\xi,\coh^u_\phi\rangle =\exp\left(\frac12 \{\phi,\xi\}\right) .
  \label{eq:cohiptl}
\end{equation}
We remark that the simple semiclassical interpretation of the coherent states as approximating classical phase space elements is lost in the evanescent case, see \cite{CoOe:locgenvac} for further discussion.

As shown in \cite{CoOe:locgenvac}, also the action of observables admits a canonical generalization from the setting of (canonical) Kähler quantization to the setting of $\alpha$-Kähler quantization. In the present case, the phase space where the observables live is now the space $L$ (or its complexification) of germs of solution on the timelike hyperplane $\Sigma$. This is no longer in correspondence to the space of initial data on a spacelike hypersurface. In particular, it includes the evanescent modes on which observables may now also depend. In analogy to Section~\ref{sec:slquant}, let $\xi\in L^\bC$ and define the linear observable $D_\xi:L^\bC\to\bC$ and the Weyl observable $F_\xi:L^\bC\to\bC$ by,
\begin{equation}
  D_\xi(\phi)\defeq 2\omega(\xi,\phi), \qquad
  F_\xi\defeq \exp(\im D) .
\end{equation}
It turns out that the formula that describes the action of the Weyl observable $\hat{F}_\xi$ on a coherent state $\coh_\phi$ takes the same form (\ref{eq:weylactcoh}) as in the Kähler quantization case, except that we have to replace complex conjugation with the real structure $\alpha_u$ \cite{CoOe:locgenvac},
\begin{equation}
  \hat{F}_\xi \coh_\phi=\exp\left(-\frac12 \{\phi,\xi\}-\frac14 \{\xi,\xi\}\right) \coh_{\phi+\xi^- +\alpha_u(\xi^-)} .
  \label{eq:weylactcoht}
\end{equation}
Note that apart from this explicit difference to formula (\ref{eq:weylactcoh}), there are also implicit differences such as the fact that $\phi$ lives here in the space $L^u$ rather than $L$.
On the other hand, the Weyl relations appear unchanged compared to (\ref{eq:weylrel}),
\begin{equation}
  \hat{F_\xi} \hat{F_\phi}=\exp(\im\omega(\xi,\phi))\hat{F}_{\xi+\phi} .
\end{equation}
The coincidence with expression (\ref{eq:weylrel}) is not unexpected, given that the relations must not depend on the representation, parametrized here by $u$. However, we emphasize that the relations we have here on a timelike hypersurface do not arise by any postulate of a formal equality to the well-known spacelike case (\ref{eq:weylrel}). Rather, both are derived as arising in the algebra of slice observables from the path integral, independent of the causal structure \cite{CoOe:locgenvac}.
The action (\ref{eq:weylactcoht}) of $\hat{F}_\xi$ is no longer unitary if $\xi$ is real. Rather, it is unitary if $\xi$ is real in the twisted sense of the real structure $\alpha_u$, i.e., if $\xi\in L^u$. Then we get on normalized coherent states,
\begin{equation}
  \hat{F_\xi} \ncoh_\phi=\exp\left(\im\omega(\xi,\phi)\right) \ncoh_{\phi+\xi} .
\end{equation}
This, again, resembles its analog (\ref{eq:weylactunitary}) in the spacelike case. Note in particular that even though $\xi$ and $\phi$ are not real we have $\omega(\xi,\phi)$ real, as follows from relation (\ref{eq:alphacomp}).

If we restrict to observables that only depend on the subspace $L^{\text{p},\bC}\subseteq L^{\bC}$ of propagating modes, i.e., $\xi\in L^{\text{p},\bC}$, there is no dependence on $u$, and the representation of the observable algebra is completely analogous to the spacelike case. In the general case we do have a dependence on $u$. In particular, as one can read off from (\ref{eq:Les}), $L^u$ and $L^{u'}$ are different subspaces of $L^{\bC}$ if $u\neq u'$. This implies in turn that even the subalgebra of the observable algebra that acts unitarily depends on $u$. Clearly, for different values of $u$ we can \emph{not} have a notion of \emph{unitary equivalence} of representations in the conventional sense. However, in apparent contradiction to conventional wisdom, this does not mean that different values of $u$ correspond to physically different theories. In particular, correlation function of observables do not depend on choices of $u$ (or more generally of $\alpha$). This is because in the present framework, correlation functions are primary objects while Hilbert spaces of states (both obtained by canonical as well as by $\alpha$-Kähler quantization) are secondary objects and by construction compatible with the former  \cite{CoOe:locgenvac}.

We proceed to comment on the relation to previous work. The construction of the Hilbert space $\cH^{\text{p}}$ of the propagating germs is equivalent to the construction given in \cite{Oe:timelike} in terms of the Schrödinger representation. It is also equivalent to the construction given in \cite{Oe:holomorphic} in terms of the holomorphic representation. The latter is easy to see by merely verifying coincidence of the complex structure $J^{\text{p}}:L^{\text{p},\bC}\to L^{\text{p},\bC}$. With the conventions of the present section this takes the form,
\begin{equation}
  (J^{\text{p}}(\phi))^{\text{a}}(E,\tilde{k})
  =-\im \phi^{\text{a}}(E,\tilde{k}) \qquad
  (J^{\text{p}}(\phi))^{\text{b}}(E,\tilde{k})
  =\im \phi^{\text{b}}(E,\tilde{k}) .
\end{equation}
As we have seen, the physically correct vacuum for evanescent modes does not correspond to a Kähler polarization and thus is not amenable for the application of canonical quantization. This problem was addressed in \cite{Oe:holomorphic} by proposing an ad-hoc Kähler polarization in the form of a complex structure. However, as follows from the results of \cite{CoOe:vaclag,CoOe:locgenvac}, such a construction will in general not lead to correct correlation functions, as it ignores the physical vacuum. However, it was shown in \cite[Section~7.6]{CoOe:locgenvac} that under particular circumstances, such a Kähler quantization can still be “correct” as long as only amplitudes and their compositions are considered and not general correlation functions. It turns out that we are precisely in this situation here. The key ingredient is a complex linear bijection $I:L^{\bC}\to L^{\bC}$, which restricts to an identification of real subspaces $L\to L^\alpha$. As shown in \cite[Section~7.7]{CoOe:locgenvac}, a positive-definite reflection map necessarily gives rise to such a linear bijection in the form $I(\phi)=\frac{1}{\sqrt{2}}(\phi+\alpha(\overline{\phi}))=\frac{1}{\sqrt{2}}(\phi-\im\gamma(\phi))$. Here, this is $I^{\text{e},u}:L^{\text{e},\bC}\to L^{\text{e},\bC}$ with explicit form,
\begin{align}
  (I^{\text{e},u}(\phi))^{\text{x}}(E,\tilde{k})
  & = \frac{1}{\sqrt{2}} \phi^{\text{x}}(E,\tilde{k})
      -\frac{\im}{\sqrt{2}} e^{2 k_1 u} \phi^{\text{i}}(E,\tilde{k}), \\
  (I^{\text{e},u}(\phi))^{\text{i}}(E,\tilde{k})
  & = \frac{1}{\sqrt{2}} \phi^{\text{i}}(E,\tilde{k})
      -\frac{\im}{\sqrt{2}} e^{-2 k_1 u} \phi^{\text{x}}(E,\tilde{k}) .
\end{align}
This then defines in particular a Kähler polarization with complex structure $\tilde{J}^{\text{e},u}=\im\gamma^u \circ J^{\text{e}}$, where $J^{\text{e}}$ is the complex structure on $L^{\text{e},\bC}$ that has eigenvalues $\im$ and $-\im$ respectively on $L^{\text{e},+}$ and $L^{\text{e},-}$. Here,
\begin{equation}
  (\tilde{J}^{\text{e},u}(\phi))^{\text{x}}(E,\tilde{k})
  =-e^{2 k_1 u} \phi^{\text{i}}(E,\tilde{k}),\qquad
  (\tilde{J}^{\text{e},u}(\phi))^{\text{i}}(E,\tilde{k})
  =e^{-2 k_1 u}\phi^{\text{x}}(E,\tilde{k}) .
  \label{eq:csetl}
\end{equation}
With this complex structure we can perform an ordinary canonical quantization. Let us call the Hilbert space obtained in this way, $\tilde{\cH}^{\text{e},u}$. This Hilbert space is then equivalent to $\cH^{\text{e},u}$ by a unitary map $\tilde{U}^{\text{e},u}:\tilde{\cH}^{\text{e},u}\to \cH^{\text{e},u}$ \cite[Section~7.6]{CoOe:locgenvac}. On coherent states $\tilde{U}^{\text{e},u}$ takes the simple form, 
$\tilde{U}^{\text{e},u}(\tilde{\coh}_\xi)=\coh_{I^{\text{e},u}(\xi)}$ for $\xi\in L^{\text{e}}$.
What is more, this canonical quantization in itself yields correct amplitudes and compositions (as long as no observables are inserted), intertwined precisely by the map $\tilde{U}^{\text{e},u}$. It turns out that the complex structure proposed in \cite{Oe:holomorphic} for the evanescent modes is precisely the one of equations (\ref{eq:csetl}) when setting $u=0$. So even though the underlying physical principle was lacking there, the proposed ad-hoc quantization was at least correct in amplitudes and compositions. (Observables were beyond the scope of that paper.)

Finally, for later use we exhibit the decomposition of the space of evanescent germs into “position” and “momenta” subspaces, $L_z=N_z\oplus M_z$, separately for propagating and evanescent modes. These subspaces can be read off from the symplectic potential (\ref{eq:tlsympote}) as null subspaces to the left and right respectively,
\begin{align}
  N_z^{\text{p},\bC} & =\{\phi\in L_z^{\text{p},\bC} : [\phi,\eta]^{\text{p}}_z=0\, \forall \eta\in L_z^{\text{p},\bC}\}=\{\phi\in L_z^{\text{p},\bC} : \phi^{\text{b}}(E,\tilde{k})=e^{2\im k_1 z}\phi^{\text{a}}(E,\tilde{k})\}, \\
  M_z^{\text{p},\bC} & =\{\phi\in L_z^{\text{p},\bC} : [\eta,\phi]^{\text{p}}_z=0\, \forall \eta\in L_z^{\text{p},\bC}\}=\{\phi\in L_z^{\text{p},\bC} : \phi^{\text{b}}(E,\tilde{k})=-e^{2\im k_1 z}\phi^{\text{a}}(E,\tilde{k})\} ,\label{eq:msprop} \\
  N_z^{\text{e},\bC} & =\{\phi\in L_z^{\text{e},\bC} : [\phi,\eta]^{\text{e}}_z=0\, \forall \eta\in L_z^{\text{e},\bC}\}=\{\phi\in L_z^{\text{e},\bC} : \phi^{\text{x}}(E,\tilde{k})=e^{2k_1 z}\phi^{\text{i}}(E,\tilde{k})\}, \\
  M_z^{\text{e},\bC} & =\{\phi\in L_z^{\text{e},\bC} : [\eta,\phi]^{\text{e}}_z=0\, \forall \eta\in L_z^{\text{e},\bC}\}=\{\phi\in L_z^{\text{e},\bC} : \phi^{\text{x}}(E,\tilde{k})=-e^{2k_1 z}\phi^{\text{i}}(E,\tilde{k})\} . \label{eq:mseva}
\end{align}
These subspaces form a pair of transversal Lagrangian subspaces, both in the propagating as in the evanescent case.

\section{Spatial evolution}
\label{sec:spevol}

\begin{figure}
  \centering
  \begin{tikzpicture}
\filldraw[gray!10] (1,-1) rectangle (3,3);
\draw[->] (-1,0) -- (5,0) node [right] {$x_1$};
\draw[->] (0,-1) -- (0,3) node [left] {$t$};
\draw[-] (1,-1) -- (1,3);
\draw[-] (3,-1) -- (3,3);
\node at (1,0) [below left] {$z$};
\node at (3,0) [below right] {$z'$};
\draw[very thick,->] (1.2,1.5) -- (2.8,1.5);
\end{tikzpicture}
  \caption{Spatial evolution through the region between an initial (left) timelike hyperplane at $x_1=z$ and a final (right) timeliek hypersurface at $x_1=z'$.}
  \label{fig:spevol}
\end{figure}
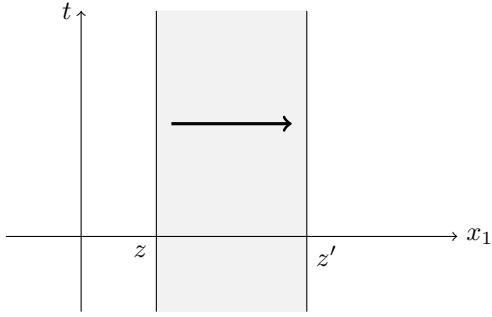

We proceed to consider the amplitude for a region bounded by pairs of timelike hyperplanes, located say at $x_1=z$ and $x_1=z'$ with $z'>z$, see Figure~\ref{fig:spevol}. We denote this region by $[z,z']$. Such a “spatial transition amplitudes” encodes a “spatial evolution". The direction of this spatial evolution is of course an arbitrary choice we can make, say we consider evolution from left to right, that is from $z$ to $z'$. We take the Hilbert spaces of states associated to the hyperplanes to be constructed precisely as specified in the previous section. In particular, we take the compatible real structures $\alpha$, $\alpha'$ as induced by reflections at hyperplanes located at $x_1=u$ and $x_1=u'$ respectively. Correspondingly we use the notation $\cH^u$, $\cH^{u'}$ of the previous section for these Hilbert spaces, but add the subscript $z$ or $z'$ when useful for indicating the associated hypersurface. Also, an over-line on the subscript indicates opposite orientation of the hypersurface.

We denote the amplitude map for the region by $\rho_{[z,z']}:\cH_{z}^{u}\tens\cH_{\overline{z'}}^{u'}\to\bC$. Note that the hypersurface at $z'$ carries opposite orientation compared to the one at $z$ as a boundary of the region $[z,z']$. Recall that there is a complex conjugate-linear involution relating the Hilbert spaces associated to the same hypersurface, but with opposite orientation, here $\iota_{z'}:\cH_{z'}^{u'}\to\cH_{\overline{z'}}^{u'}$. We denote the induced spatial evolution map by $V_{[z,z']}^{u,u'}:\cH_{z}^{u}\to\cH_{z'}^{u'}$. Its relation to the amplitude map is,
\begin{equation}
  \langle \psi', V_{[z,z']}^{u,u'}\psi\rangle_{z'}=\rho_{[z,z']}(\psi\tens\iota_{z'}(\psi')) .
\end{equation}
For coherent states we have a closed formula for the amplitude \cite{CoOe:locgenvac}. In the case at hand this turns out to be particularly simple due to the space-interval structure of the region. Take $\xi\in L^u_z$ and $\xi'\in L^{u'}_{z'}$ to label coherent states $\coh^{u}_{z,\xi}\in\cH^u_z$ and $\coh^{u'}_{z',\xi'}\in\cH^{u'}_{z'}$. Then,
\begin{equation}
  \langle \coh^{u'}_{z',\xi'}, V_{[z,z']}^{u,u'}\coh^{u}_{z,\xi}\rangle_{z'}
  =\rho_{[z,z']}(\coh_{z,\xi}^{u}\tens\coh_{\overline{z'},\xi'}^{u'})
  =\exp\left(\frac12 \{\xi,\xi'\}\right) .
  \label{eq:cohevolampl}
\end{equation}
Again, hypersurface locations do not play an explicit role, and we omit the corresponding labels in the following. The right-hand side of this expression takes exactly the same form as in the inner product (\ref{eq:cohiptl}) of coherent states, but there is an important difference. In contrast to the inner product formula, the elements $\xi$ and $\xi'$ here live in different spaces. It turns out that we are dealing nevertheless with an inner product of coherent states, i.e., that $V^{u,u'}\coh^{u}_{\xi}$ is a coherent state. To see this we suppose that there is a linear map $v^{u,u'}: L^u\to L^{u'}$ so that,
\begin{equation}
  V^{u,u'}\coh^{u}_{\xi}= \coh^{u'}_{v^{u,u'}(\xi)} .
  \label{eq:cohevol}
\end{equation}
Then, we would have, by the inner product formula (\ref{eq:cohiptl}),
\begin{equation}
  \langle \coh^{u'}_{\xi'}, V^{u,u'}\coh^{u}_{\xi}\rangle
  =\langle \coh^{u'}_{\xi'}, \coh^{u'}_{v^{u,u'}(\xi)}\rangle
  =\exp\left(\frac12 \{v^{u,u'}(\xi),\xi'\}\right) .
  \label{eq:cohevolip}
\end{equation}
Comparison of formulas (\ref{eq:cohevolampl}) and (\ref{eq:cohevolip}) yields the condition
\begin{equation}
  (v^{u,u'}(\xi))^{-}=\xi^{-} ,
\end{equation}
which has a unique solution. This confirms the validity of formula (\ref{eq:cohevol}). What is more, $v^{u,u'}$ extends uniquely to a complex linear map $L^{\bC}\to L^{\bC}$,
\begin{equation}
  v^{u,u'}(\xi)=\xi^{-}+\alpha_{u'}(\alpha_{u}(\xi^+))
  =\xi^{-}+\gamma^{\text{e}}_{u'}(\gamma^{\text{e}}_{u}(\xi^{+})) .
\end{equation}
We use the notation $\gamma_u^{\text{e}}$ here for the map that acts as the reflection $\gamma_u$ on $L^{\bC,\text{e}}$, given by (\ref{eq:reflects}), and as the identity on $L^{\bC,\text{p}}$.
We can read off that the map $v^{u,u'}$ leaves invariant the evanescent solutions that decay to the right and moves the solutions that decay to the left by an amount $2(u'-u)$ to the right.

In terms of the global parametrization of solutions that we have chosen, the maps $v^{u,u'}$ and $V^{u,u'}$ are trivial on propagating solutions and in particular unitary. This is completely analogous to the temporal evolution of states on spacelike hypersurfaces (Section~\ref{sec:slquant}). For evanescent solutions this is true only if we have chosen a quantization based on the same parameter $u$ for “initial” (left) and “final” (right) hyperplane. In the case $u\neq u'$ the map $v^{u,u'}$ from $L^u$ to $L^{u'}$ is not unitary with respect to the inner product (\ref{eq:rjipe}). Explicitly, we have,
\begin{equation}
  \{v^{u,u'}(\phi),v^{u,u'}(\eta)\}
  =\{\phi,\gamma^{\text{e}}_{u'}(\gamma^{\text{e}}_{u}(\eta))\}
  =2\im\int_{|E|< E_\parallel}
  \frac{\xd^2\tilde{k}\,\xd E}{(2\pi)^3 2k_1}
  e^{2 k_1 (u-u')}
  \eta^\text{i}(-E,-\tilde{k})\phi^\text{x}(E,\tilde{k}) ,
\end{equation}
compare to (\ref{eq:rjipe}). Consequently, the evolution map $V^{u,u'}$ between the corresponding Hilbert spaces is not unitary either. This follows from the fact that $V^{u,u'}$ is really the second quantization of $v^{u,u'}$. More explicitly, we can read this off from (\ref{eq:cohevol}), noting that the norm of a coherent state in $\cH^u$ is determined via (\ref{eq:cohiptl}) and thus ultimately again in terms of the inner product (\ref{eq:rjipe}) on $L^u$.

As far as the action of the observable algebra(s) on the Hilbert spaces for timelike hyperplanes at different positions, the evolution map $V^{u,u'}$ is an intertwiner. Again, due to our global parametrization, the positions of the hyperplanes do not matter, but only the quantization parameters $u,u'$. In particular, for the Weyl observables as defined in Section~\ref{sec:Hstml} we have,
\begin{equation}
  \hat{F}_\xi V^{u,u'}=V^{u,u'} \hat{F}_\xi .
  \label{eq:obsinter}
\end{equation}
Note that we use the same notation $\hat{F}_\xi$ for operators on different Hilbert spaces here. What is more, the generalization of equation (\ref{eq:obsinter}) to arbitrary observables only holds for those that are well-defined on both Hilbert spaces $\cH^{u}$ and $\cH^{u'}$. While the intertwining property can be easily demonstrated from the formulas we have already provided, this is not really necessary. Rather, this property follows from the fact that both sides of equation (\ref{eq:obsinter}) represent just different “translations” of the same object in the formalism based on correlation functions and the path integral that underlies our considerations \cite{CoOe:locgenvac}.

\section{Probability of spatial transition}
\label{sec:transprob}

In the conventional wisdom of the standard formulation of quantum theory, unitarity of evolution in time appears as an indispensable ingredient of a consistent probability interpretation. The spatial evolution between timelike hypersurfaces we are discussing here is quite analogous to the temporal evolution between spacelike hypersurfaces in that we have a one-to-one correspondence between data on the “initial” and “final” hypersurfaces. The reader might thus be worried that the lack of unitarity of the spatial evolution maps $V^{u,u'}$ (if $u\neq u'$) spells trouble for the consistency of the theory and might limit us to choose the same parameter $u$ for all timelike hyperplanes. However, this is not the case. The general boundary formulation (GBF) or positive formalism in which we are working does not require unitarity (even for temporal evolution) to ensure a consistent probability interpretation. Rather, positivity combined with compositionality do that \cite{Oe:posfound}. A price to pay is that probability formulas look (slightly) more complicated. The simpler formulas of the standard formulation arise as a special case when additional conditions are met (such as unitarity) and certain normalizations are imposed \cite{Oe:posfound}.

We wish to consider “spatial transition” probabilities in analogy to the standard transition probabilities associated with evolution in time.
However, to separate the issue of non-unitarity from the issue of spatial vs.\ temporal evolution we concentrate on the former first. Thus, we suppose that we have initial and final Hilbert spaces $\cH$ and $\cH'$ at times $t$ and $t'$, with $t'> t$. Furthermore, evolution is described by a not necessarily unitary map $E:\cH\to\cH'$. Given a normalized initial state $\psi$ and a normalized final state $\psi'$, what is the probability to measure $\psi'$ at $t'$, given that $\psi$ was prepared at time $t$? In general, the probability for a measurement on the boundary $\partial M$ of a spacetime region $M$ may be obtained as follows. In the \emph{boundary Hilbert space} $\cH_{\partial M}$ we consider two \emph{positive operators}, $\cS,\cA\in \op^+(\cH_{\partial M})$. (Here $\op^+(\cH_{\partial M})$ denotes the set of positive operators on $\cH_{\partial M}$.) $\cS$ encodes the \emph{preparation} and any knowledge we have about the measurement. $\cA$ encodes additionally an affirmative answer to the \emph{question} we pose. $\cA$ represents a situation more special than that corresponding to $\cS$, which translates to the inequality $\cA\le\cS$. Recall that the positive \emph{probability map} $A_M:\op^+(\cH_{\partial M})\to\R^+$ is the mixed-state analogue of the \emph{amplitude map} $\rho_M:\cH_{\partial M}\to\bC$ \cite{Oe:dmf}. The \emph{probability} $P$ for an affirmative answer to the measurement question is given by the quotient,
\begin{equation}
  P=\frac{A_M(\cA)}{A_M(\cS)} .
\end{equation}
In the case at hand, $M$ is the time interval $[t,t']$, $\cH_{\partial M}=\cH\tens{\cH'}^*$, $\cS=\po_\psi\tens\id$ and $\cA=\po_\psi\tens \po_{{\psi'}^*}$ \cite{Oe:gbqft}. The result is,
\begin{equation}
  A_M(\cA)=|\langle \psi', E\psi\rangle |^2,\qquad 
  A_M(\cS)=\langle E\psi, E\psi\rangle,\qquad
  P=\frac{|\langle \psi', E\psi\rangle |^2}{\langle E\psi, E\psi\rangle} .
  \label{eq:tprob}
\end{equation}
Crucially, this makes sense even if $E$ is not a unitary map. Alternatively, we could have tried to guess this formula by starting from the usual formula of the standard formulation, which is just the numerator. Then, modifying this by hand to satisfy obvious consistency conditions could have led us to the right answer. Of course, it would then have been unclear how to generalize this to other types of measurements and whether this could be made consistent under composition of measurements. Fortunately, with the consistent framework of the GBF and the positive formalism \cite{Oe:posfound} at our disposal we have the luxury of no longer needing to worry about this.

Mathematically, the change from a temporal transition probability to spatial transition probability is trivial. We now ask, given a state $\psi$ at $z$, what is the probability of having a state $\psi'$ at $z'$? The answer takes the same form (\ref{eq:tprob}), with the difference that the Hilbert spaces $\cH$ and $\cH'$ are now associated to timelike hypersurfaces and $E$ encodes spatial evolution. To be more concrete, consider normalized coherent states $\ncoh_\xi^u\in\cH_z^u$, $\ncoh_{\xi'}^{u'}\in\cH_{z'}^{u'}$. The probability of finding $\ncoh_{\xi'}^{u'}$ at $z'$, given that we have $\ncoh_{\xi}^{u}$ at $z$ is thus,
\begin{equation}
  P=\frac{|\langle\ncoh_{\xi'}^{u'}, V^{u,u'}\ncoh_{\xi}^{u}\rangle |^2}{\langle V^{u,u'}\ncoh_{\xi}^{u}, V^{u,u'}\ncoh_{\xi}^{u}\rangle}
  =\exp\left(-\frac12\{\xi'-v^{u,u'}(\xi),\xi'-v^{u,u'}(\xi)\}\right) .
\end{equation}
Since the inner product $\{\cdot,\cdot\}$ is positive-definite on $L^{u'}$, we obtain in the argument of the exponential a non-positive real number. In turn, the exponential yields a value between $0$ and $1$ (although $0$ cannot be attained). As expected, if the image of the “initial” (left) coherent state equals (up to normalization) the “final” (right) coherent state, the probability is unity. It decreases exponentially with increasing difference, measured in the “phase space” $L^{u'}$. This result is exactly what we should expect physically. Crucially, however, we have to be aware that this probability is something quite different from the usual measurement probabilities. Here, we are not preparing a state and then making a measurement. Rather, the state on the left (at $z$) is something that persists over all time and that we nevertheless pretend to know for certain. On the other hand, the answer to the question about the state on the right (at $z'$) is not something that is revealed by a measurement at a specific instant of time. Rather, it pertains to all times. For relevant previous discussion of measurements on timelike hypersurfaces, see \cite{Oe:timelike,Oe:kgtl,CoOe:smatrixgbf}.
Another caveat concerns the interpretation of the coherent states. In Kähler quantization the coherent states are labeled by elements of the corresponding real phase space $L$ and admit an interpretation as quantum analogs of the corresponding classical phase space element. This is true in the present context for the propagating modes. The evanescent modes on the other hand are dealt with via $\alpha$-Kähler quantization. Consequently, their coherent states are labeled by elements of $L^\alpha$, which generally are not elements of the real phase space. A semiclassical interpretation for these coherent states is more complicated \cite{CoOe:locgenvac}.

% !TEX root = main.tex

\section{The Casimir state}
\label{sec:casimir}

Recall that the Casimir effect is the occurrence of an attractive electromagnetic force between parallel conducting plates that originates from the alteration of the quantum vacuum between the plates due to the conducting boundary conditions \cite{Cas:effect}. From the present perspective, the conducting boundary condition literally amounts to an alternative vacuum in terms of an alternative Lagrangian subspace on the timelike hypersurface given by the conducting plate \cite{CoOe:locgenvac}. Moreover, it gives rise to a pseudo-state that can be interpreted, as we shall see, as a sea of particle pairs representing this conducting vacuum as seen from the standard vacuum. We call this state the \emph{Casimir state}. Note that this is distinct from the particle production originating from the acceleration of conducting plates (also known as “moving mirrors”) which has been the subject of investigation in the literature for a long time \cite{Dew:qftcurved}. In contrast, we show here that even in the much simpler static case, a sea of virtual particle pairs is present near the \emph{timelike} hyperplane which is the trajectory of the conducting plate. We work in the Klein-Gordon theory, building on the quantization discussed in the previous sections.

Recall from Section~\ref{sec:Hstml} that the standard (left) vacuum on the timelike hyperplane located at $x_1=z$ (see Figure~\ref{fig:tlhp}) is encoded in the Lagrangian subspace $L_z^+\subseteq L_z^{\bC}$ given by (\ref{eq:tlpvac}) and (\ref{eq:tlevac}). On the other hand, the conducting boundary condition is simply the Dirichlet condition imposing the solution to vanish on the hypersurface. The corresponding subspace is thus $M_z^\bC\subseteq L_z^\bC$, decomposing into propagating and evanescent components, $M_z^\bC=M_z^{\text{p},\bC} \oplus M_z^{\text{e},\bC}$ given by (\ref{eq:msprop}) and (\ref{eq:mseva}). It follows from the results of \cite{CoOe:locgenvac} that there exists a pseudo-state $Y_z^{\text{Cas}}$, the Casimir state, in the Hilbert space $\cH_z^u$ that encodes the generalized vacuum given by the Dirichlet boundary condition. To be precise, $Y_z^{\text{Cas}}$ is not a state in $\cH_z^u$ as it is not normalizable. Nevertheless, $Y_z^{\text{Cas}}$ is perfectly well-defined as a $J$-holomorphic wave function on $L_z^u$ in the holomorphic representation, but it is not square-integrable. This is quite analogous to the well-known pseudo-states induced by inequivalent vacua in the standard Kähler quantization on spacelike hypersurfaces. What is more, the wave function extends to an (ordinary) holomorphic function on $L^{\bC}$, which in the present context of $\alpha$-Kähler quantization does not depend on the real structure $\alpha$ (here encoded by $u$). We recall the formula \cite[Section~7.4]{CoOe:locgenvac},
\begin{equation}
  Y_z^{\text{Cas}}(\xi)=\exp\left(\frac14\{\xi^{\mathrm{X}},\xi\}_z\right) ,
\end{equation}
where we decompose $\xi=\xi^{\text{X}}+\xi^{\text{M}}$ according to $L_z^{\bC}=M_z^{\bC}\oplus L_z^{-}$. This yields,
\begin{equation}
  Y_z^{\text{Cas}}(\xi)
  =\exp\left(-\frac{1}{2}\int\frac{\xd^2\tilde{k}\,\xd E}{(2\pi)^3 2k_1}\left(e^{-2\im k_1 z}
  \xi^{\text{b}}(E,\tilde{k})\xi^{\text{b}}(-E,-\tilde{k})
  +\im e^{2 k_1 z}
  \xi^{\text{i}}(E,\tilde{k})\xi^{\text{i}}(-E,-\tilde{k})\right)
  \right) .
  \label{eq:caswf}
\end{equation}
Here we have written the propagating and evanescent contributions under the same integral with the understanding that we set $\xi^{\text{b}}(E,\tilde{k})=0$ if $|E|< E_\parallel$ and $\xi^{\text{i}}(E,\tilde{k})=0$ if $|E|> E_\parallel$.

In order to facilitate a particle interpretation we introduce the following notation for plane wave solutions and exponential solutions in $L_z^{\bC}$,
\begin{equation}
  \varphi_{E,\tilde{k}}^{\text{p}}(t,x_1,\tilde{x}) =e^{-\im(E t -\tilde{k}\tilde{x} +k_1 (x_1-z))}, \qquad
  \varphi_{E,\tilde{k}}^{\text{e}}(t,x_1,\tilde{x})=e^{-\im\pi/4}e^{-\im(E t -\tilde{k}\tilde{x}) +k_1 (x_1-z)} . \label{eq:tlsols}
\end{equation}
Next, we consider the induced real solutions in $L_z^u$,
\begin{align}
  \tilde{\varphi}_{E,\tilde{k}}^{\text{p}}(t,x_1,\tilde{x})
  & =\Re(\varphi_{E,\tilde{k}}^{\text{p}})
     =\cos(E t -\tilde{k}\tilde{x} +k_1 (x_1-z))
  , \nonumber \\
  \tilde{\varphi}_{E,\tilde{k}}^{\text{e}}(t,x_1,\tilde{x})
  & =P^{\text{e},u}(\varphi_{E,\tilde{k}}^{\text{e}})
   =e^{- \im\pi/4} \cos(Et-\tilde{k}\tilde{x}+\im k_1(x_1-u)) e^{k_1(u-z)} 
  .
\end{align}
Here, $P^{\text{e},u}=\frac12 (1+\alpha^{\text{e}}_u)$ is the projector onto $L^{\text{e},u}$. 
Using the collective notation $\tilde{\varphi}_{E,\tilde{k}}$ for either $\tilde{\varphi}_{E,\tilde{k}}^{\text{p}}$ or $\tilde{\varphi}_{E,\tilde{k}}^{\text{e}}$, depending on whether $|E|> E_{\parallel}$ or $|E|< E_{\parallel}$ we rewrite (\ref{eq:caswf}),
\begin{equation}
  Y_z^{\text{Cas}}(\xi)
  =\exp\left(-\frac{1}{2}\int\frac{\xd^2\tilde{k}\,\xd E}{(2\pi)^3 2k_1}
    \{\tilde{\varphi}_{E,\tilde{k}},\xi\}_z
    \{\tilde{\varphi}_{-E,-\tilde{k}},\xi\}_z
  \right) .
\end{equation}
In terms of creation operators, this is, with the obvious notation,
\begin{equation}
  Y_z^{\text{Cas}}
  =\exp\left(\int\frac{\xd^2\tilde{k}\,\xd E}{(2\pi)^3 2k_1}
    a^\dagger_{E,\tilde{k}} a^\dagger_{-E,-\tilde{k}}
  \right) \coh^u_{z,0} .
\end{equation}
The Casimir state consists of a sea of pairs of particles. The two particles comprising each pair have opposite quantum numbers, here given in terms of the “energy” $E$ and the 2-momentum tangential to the hypersurface $\tilde{k}$. Crucially, in our present parametrization negative values of $E$ do not encode physically negative energies. Rather, the sign of $E$ encodes in the \emph{propagating} case the binary degree of freedom distinguishing \emph{in-coming} from \emph{out-going} particles \cite{Oe:timelike,Oe:kgtl}. Here, in-coming or out-going is meant with respect to the region to the right of the hypersurface $\Sigma$, i.e.,  with $x_1>z$. As can be read off from (\ref{eq:tlsols}), $E>0$ corresponds to out-going, that is left-moving particles, while $E<0$ corresponds to in-coming, i.e., right-moving particles. Intuitively, we may express the situation as follows: We have a sea of particle pairs, with one particle in each pair emitted from the conducting plate, while the other is absorbed by it.

Apart from the usual propagating particles (although in the unusual context of a timelike hypersurface) we have the novel phenomenon of \emph{evanescent particles}. As the propagating ones, the conducting plate creates these in pairs. Naturally, however, there is no notion of these particles moving to the left or right as they correspond to exponential rather than oscillating solutions. In terms of degrees of freedom, the binary degree of the sign of the “energy” $E$ now is in correspondence to the choice between solutions decaying exponentially to the left or to the right. This does not mean, however, that in our present parametrization a definite sign of the energy corresponds to either decaying or increasing modes. To disentangle exactly the correspondence between the two manifestations of this binary degree of freedom requires a further analysis of the correspondence between classical and quantum degrees of freedom in $\alpha$-Kähler quantization, which is beyond the scope of the present work.

% !TEX root = main.tex

\section{Conclusions and Outlook}
\label{sec:conclusions}

In the present work we have presented a construction of a Hilbert space of states for Klein-Gordon theory on the timelike hyperplane. To this end we have employed the novel $\alpha$-Kähler quantization prescription introduced in \cite{CoOe:locgenvac}. This was necessary, because only the propagating modes on the hyperplane can be quantized via ordinary canonical (Kähler) quantization \cite{Oe:timelike}. As shown in \cite{CoOe:vaclag}, the evanescent modes, which are also present on the hypersurface, have decaying asymptotic boundary conditions at spatial infinity which induces a vacuum encoded in a real polarization on the phase space. $\alpha$-Kähler quantization is precisely designed for this situation and involves a compatible real structure $\alpha$, which gives rise to a twisted $*$-structure on the quantum algebra of slice observables on the hypersurface in question. As we have seen in Section~\ref{sec:Hstml}, a suitable real structure $\alpha$ can be obtained from a reflection map $\gamma_u$. The latter encodes reflection at a timelike hyperplane situated at a fixed value $u$ of the coordinate $x_1$. As explained in  \cite{CoOe:locgenvac}, the use of such a reflection map is inspired by a similar construction that encodes reflection-positivity in the Euclidean approach to quantum field theory. This means that even with the vacuum fixed, we have for the evanescent modes a quantization ambiguity parametrized by this coordinate value $u\in\R$. As we have seen, different values of $u$ lead not only to inequivalent representations of observables, but the very notion of what constitutes a “real” observable is twisted and depends on $u$. Nevertheless, correlation functions of observables, and thus physical properties encoded in them, are not affected by this ambiguity as they are independent of $u$. This is by design of $\alpha$-Kähler quantization \cite{CoOe:locgenvac}.

We have investigated in Section~\ref{sec:spevol} the \emph{spatial evolution} of states in the coordinate direction $x_1$ in terms of a map from the state space on an initial timelike hyperplane to a final one. This is quite analogous to temporal evolution between spacelike hypersurfaces at different times. Indeed, for states encoding propagating modes it was already shown in  \cite{Oe:timelike} that an analogous unitary evolution results. Here, we have seen that for evanescent modes a unitary evolution is also obtained if the parameter $u$ is chosen the same for the Hilbert spaces of both hyperplanes. However, if the parameter is chosen differently and the quantizations are hence inequivalent, the evolution is not unitary. As laid out in Section~\ref{sec:transprob}, this does not impede a well-defined probability interpretation for spatial transitions between states. Indeed, using the underlying positive formalism \cite{Oe:posfound} we were able to show explicitly how such transition probabilities are calculated consistently.

In its simplest manifestation the Casimir effect arises from conducting boundary conditions on a pair of timelike hyperplanes. Taking a single hyperplane, the insights obtained in \cite{CoOe:vaclag} allow us to interpret this boundary condition as an alternative vacuum. With the methods developed in \cite{CoOe:locgenvac} we can represent this as a pseudo-state on the timelike hyperplane. This pseudo-state is not normalizable in the Hilbert space of states of the standard vacuum, but is well-defined as a holomorphic wave function. We have constructed this wave function is Section~\ref{sec:casimir} and shown that it corresponds to sea of particle pairs with opposite quantum numbers. This sea of particle pairs includes propagating particles which are emitted from and absorbed into the hyperplane as well as evanescent particles near the hyperplane. In this way, the origin of the Casimir effect turns out to be quite analogous to other situations in quantum field theory where a modified vacuum is the root cause, such as in Hawking radiation.

We turn to a number of open questions and future research directions associated with our results. As shown, the construction of the Hilbert space for the evanescent sector on the timelike hyperplane suffers from a quantization ambiguity. This comes in the form of the choice of a compatible real structure $\alpha$ on the phase space of the timelike hyperplane. In the present paper we have derived $\alpha$ from a reflection map $\gamma$ depending on a parameter $u\in\R$ denoting the location of the reflecting hyperplane. In general one may ask what a good choice for $\alpha$ is or how a classification of possible choices looks like. One selection criterion that comes to mind would be the invariance under symmetries, particularly spacetime symmetries. In the present case, at least the $\alpha$ obtained from reflection at a hypersurface is invariant under all isometries that do not affect the $x_1$-coordinate. However, there might be other $\alpha$ with the same invariance or with other desirable properties.

A very important open question concerns the relation between classical and quantum objects. $\alpha$-Kähler quantization as exhibited here for the evanescent modes is mathematically compelling and by construction consistent with path integral quantization and quantum field theoretic correlation functions. However, it lacks the direct correspondence that canonical quantization establishes between classical observables on the phase space and hermitian operators that can be interpreted as giving rise to projection valued measurements. Instead, certain complex functions on phase space, that are real in an $\alpha$-twisted sense, give rise to hermitian operators and thus lead to a notion of measurement. But what is measured if the originating classical object is not a conventional observable? To get to the bottom of this might require a deeper understanding of how to construct generalized quantum operations (also called \emph{probes}) from classical field theory observables in the positive formalism \cite{Oe:posfound}.

An exciting possibility opened by the present work is a conceptually satisfactory approach to the time-of-arrival problem in quantum theory. This is the problem of predicting the time that a particle hits a screen (or more general detector). While the corresponding problem of predicting where a particle might be detected at a certain time is quite elementary in the standard formulation of quantum theory, the analogous temporal problem is still not completely resolved \cite{MuSMEg:timeinqm1}. However, in the same way that ordinary states on a spacelike hypersurface may be localized in space, states on a timelike hypersurface may be localized in time (and the two tangential spatial dimensions). This is why the present work might provide a crucial step to address this problem.

\subsection*{Acknowledgments}

This publication was made possible through the support of the ID\# 61466 grant from the John Templeton Foundation, as part of the “The Quantum Information Structure of Spacetime (QISS)” Project (qiss.fr). The opinions expressed in this publication are those of the authors and do not necessarily reflect the views of the John Templeton Foundation.

%\appendix

\newcommand{\eprint}[1]{\href{https://arxiv.org/abs/#1}{#1}}
\bibliographystyle{stdnodoi} % bibliography
\bibliography{stdrefsb}
\end{document}